\documentclass[twocolumn,amssymb,nofootinbib, nobibnotes,aps,prd]{revtex4}

\usepackage{graphicx}
\usepackage[dvips]{color}
\usepackage{amsmath,amsthm,amssymb}

\newcommand{\be}{\begin{equation}}
\newcommand{\ee}{\end{equation}}
\newcommand{\ba}{\begin{eqnarray}}
\newcommand{\ea}{\end{eqnarray}}

\renewcommand{\vec}[1]{\boldsymbol{#1}}

\def\ap{\approx}
\def\d{{\rm d}}

\def\lsim{\raise0.3ex\hbox{$\;<$\kern-0.75em\raise-1.1ex\hbox{$\sim\;$}}}
\def\gsim{\raise0.3ex\hbox{$\;>$\kern-0.75em\raise-1.1ex\hbox{$\sim\;$}}}

\def\theta{\vartheta}

\def\Rm{{\cal R}_{\rm max}}

\begin{document}

\title{Escape model for Galactic cosmic rays 
  and an early extragalactic transition}

\author{G.~Giacinti$^{1}$}
\author{M.~Kachelrie\ss$^{2}$}
\author{D.~V.~Semikoz$^{3}$}
\affiliation{$^{1}$University of Oxford, Clarendon Laboratory, Oxford, United Kingdom}
\affiliation{$^{2}$Institutt for fysikk, NTNU, Trondheim, Norway}
\affiliation{$^{3}$AstroParticle and Cosmology (APC), Paris, France}

\begin{abstract}
We show that the cosmic ray (CR) knee can be entirely explained by 
energy-dependent CR leakage from the Milky Way, with an excellent fit 
to all existing data. We test this hypothesis calculating the trajectories 
of individual CRs in the Galactic magnetic  field.  We find that the 
CR escape time $\tau_{\rm esc}(E)$ 
exhibits a knee-like structure around $E/Z={\rm few}\times 10^{15}$\,eV for 
small coherence lengths and strengths of the turbulent magnetic field. 
The  resulting intensities for different groups of nuclei are consistent 
with the ones determined by KASCADE and KASCADE-Grande, using simple 
power-laws as injection spectra. 
The transition from Galactic to extragalactic CRs is terminated at 
$\ap 2\times 10^{18}$\,eV, while extragalactic CRs contribute significantly
to the subdominant proton flux already for $\gsim 2\times 10^{16}$\,eV.
The natural source of  extragalactic CRs in the intermediate energy region 
up to the ankle are in this model normal and starburst galaxies. 
The escape model provides a good fit to 
$\ln(A)$ data; it predicts that the phase of the CR dipole varies strongly 
in the energy range between $1\times 10^{17}$ and $3\times 10^{18}$\,eV,
while our estimate for the dipole magnitude is consistent with observations.
\end{abstract}


\maketitle




\section{Introduction}
\label{Introduction}


The all-particle cosmic ray (CR) energy spectrum is a nearly featureless 
power-law between $\sim 10^{10}$\,eV and $\sim 10^{20}$\,eV, with only a 
few breaks in its spectral index. The two most prominent ones are the knee 
at $E_{\rm k} \approx 4$\,PeV, and the ankle at $E_{\rm a} \approx 4$\,EeV. 
They must contain information about either CR sources or CR propagation. 
The range of possible theoretical explanations for these two breaks has 
been reduced in the last decade, but there is still no firm consensus 
on their origins. Another related open question is where the transition 
from Galactic to extragalactic CRs occurs. 
Unveiling this transition energy holds precious keys to understanding particle 
acceleration in the Universe. Other features have been observed in the 
spectrum between $E_{\rm k}$ and $E_{\rm a}$, see
e.g.~\cite{Apel:2012rm,Aartsen:2013wda}, and should contain additional 
clues.

In addition to the all-particle spectrum, both measurements of the 
primary composition and upper limits on the CR dipole anisotropy are 
crucial to constrain models of the knee and of the intermediate region 
up to the ankle. The KASCADE-Grande collaboration has recently provided 
measurements of the intensities of individual groups of CR nuclei up to 
$E \approx 10^{17}$\,eV~\cite{Apel:2012rm,KG,dataKG}. Also, the IceTop 
collaboration has presented measurements of the average of the 
logarithmic mass up to 30\,PeV~\cite{top}, while the Auger collaboration
deduced the contribution of individual CR groups to the total CR flux 
above $10^{17.8}$\,eV from studies of the development of air 
showers~\cite{Aab:2014kda,Aab:2014aea}.  
In the region between the knee and the ankle, upper limits on the 
amplitude of the anisotropy have been reported, at roughly the percent 
level, by KASCADE~\cite{Antoni:2003jm}, KASCADE-Grande~\cite{Curcio}, 
and Auger~\cite{PAOaniso}. Below the knee, the dipole amplitude has been 
measured at the $\sim 10^{-3}$ level, by notably Super-Kamiokande 
(10\,TeV)~\cite{Guillian:2005wp}, MILAGRO (6\,TeV)~\cite{Abdo:2008aw}, 
EAS-TOP ($\approx 100$ and 400\,TeV)~\cite{EASTOPD}, IceCube 
(20 and 400\,TeV)~\cite{Ianisp} and IceTop (2\,PeV)~\cite{Aartsen:2012ma}.

For the knee, two main explanations currently remain possible. 
First, it may be the signature of the maximum energy to which Galactic 
CR sources can accelerate protons, see e.g.~\cite{pop}. A nearby source could 
also leave such an imprint in the spectrum~\cite{Erl}. Second, the knee 
could be caused by a change in the energy dependence of the CR diffusion 
coefficient and thence confinement time in the Galaxy~\cite{sa,hall,PaperI}, 
if the CR Larmor radius is the order of the coherence length $l_{\rm c}$ of 
the turbulent Galactic magnetic field (GMF) at $E_{\rm k}$. 
In Ref.~\cite{PaperI}, we have studied this possibility---which we 
denote as the ``escape model''---by propagating individual CRs in 
recent GMF models. This enabled us to avoid limitations from the 
diffusion approximation: While reliable analytical approximations 
for the diffusion tensor are only available in certain limiting cases,
the diffusion approximation per se is not justified at the highest energies 
studied. We showed that the escape model is viable and 
can explain the individual fluxes of CR groups as measured by KASCADE
and KASCADE-Grande. Moreover, our estimate for the dipole anisotropy in
this model was consistent within uncertainties with observations.

In this work, we extend our previous study in Ref.~\cite{PaperI} and 
formulate a model for the entire energy region between 
300\,GeV/$Z$ and the ankle. In addition to the Jansson-Farrar (JF) 
GMF model~\cite{J} used in~\cite{PaperI}, we consider the Pshirkov {\it et 
al.\/} (PTKN) model~\cite{P1,P2}. This enables us to check the 
dependence of our results on the specific GMF model. Moreover, a more 
detailed study of the Galactic CR primary composition between $E_{\rm k}$ 
and $E_{\rm a}$ is presented here and compared to observations. We show 
that any remaining heavy nuclei flux in the sub-ankle region would be 
dominated by only one or few local sources. We use limits on the iron
fraction at $\gsim 7\times 10^{17}$\,eV determined by the Auger collaboration
together with $\ln(A)$ measurements to constrain the transition energy between
Galactic and extragalactic CRs, deducing $\Rm=E_{\max, \rm Fe}/(26e)\sim 10^{17}$\,V
as the maximal rigidity $\Rm$ to which Galactic  CR sources 
are able to accelerate CRs. 
The recovery of the proton and helium spectra above $E/Z\sim 10^{16}$\,eV
in the KASCADE-Grande data is maninly explained by the specific shape of the
escape rate $\tau_{\rm esc}(E)$  discovered in~\cite{PaperI}.
We show also that observational constraints from  anisotropy limits
are compatible with the escape model. A natural extension of the
'escape model' to other normal galaxies suggests that the extragalactic flux 
in the intermediate  energy region up to the ankle is composed of CRs 
accelerated in starburst galaxies.

This article in organized as follows: In Section~\ref{GMF}, we test two 
GMF models for the regular and turbulent fields, as well as different 
strengths and coherence lengths for the turbulence. We deduce a range of
models that fit constraints from notably the B/C ratio. We then compute in 
Section~\ref{IndividualFluxes} the resulting fluxes of Galactic CR groups 
and show that they fit very well KASCADE and KASCADE-Grande measurements. In 
Section~\ref{Transition}, we discuss the transition from Galactic to 
extragalactic CRs in our model. Finally, we review in Section~\ref{Discussion}
the constraints on and consequences of the escape model, before we present 
our conclusions in Section~\ref{Conclusions}.


\section{Galactic magnetic field models, parameters for the turbulence and CR confinement in the Galaxy}
\label{GMF}


\subsection{Grammage}

An important constraint on CR propagation models comes from ratios of 
stable primaries and secondaries produced by CR interactions on gas 
in the Galactic disk. In particular, the B/C ratio has been recently 
measured by the AMS-02 experiment up to 670\,GeV/nucleon~\citep{AMS02}. 
Above $E \gtrsim 10$\,GeV/nucleon, these measurements for the B/C ratio 
are consistent with a straight power-law.

In our previous work~\cite{PaperI}, we used a fit of the grammage
performed in Ref.~\cite{Jones} using the leaky-box  formalism and
earlier data. In all cases considered in~\cite{Jones}, the grammage traversed 
by CRs at reference energies $E_0/Z=(5-15)$\,GeV was found to lie in 
the range $(9-14)$\,g/cm$^2$. In order to take advantage of the high-quality
and the large energy range of the B/C data from AMS-02, we use now these 
data to derive  the grammage traversed by CRs as function of their
energies in two simple models. In the first one, we approximate the 
fraction of the B to C intensities by
\be
 \frac{I_B}{I_C} = \frac{p_{\rm sp}\lambda_s}{\lambda_B-\lambda_C } 
 \left[ \exp\left( \frac{X}{\lambda_C}-\frac{X}{\lambda_B} \right) 
        -1 \right] \,,
\ee 
where  $\lambda_i=m_p/\sigma_i$ are the interaction lengths (in gr/cm$^2$),
$\sigma_i$ the total inelastic cross section,
$m_p$ the proton mass and $p_{\rm sp}=\sigma_{\rm sp}/\sigma_{\rm tot}$ is the 
spallation probability deduced from the cross sections given in
Ref.~\cite{Spal}. In the second approximation, we employ a fit function 
giving the B/C ratio directly as function of the grammage, following the 
approach in Refs.~\cite{Webber:2003cj,Blum:2013zsa}.
In Fig.~\ref{figX1}, the grammage derived in Ref.~\cite{Jones} using earlier 
data is shown as black cross. The grammage deduced from the AMS-02
data using the first approximation is shown with magenta error-bars,
while the grammage obtained using the second approximation is shown with 
blue error-bars. Note that the error-bars take into account only the statistical
and systematic errors of the AMS-02 measurement, while uncertainties in the
cross sections or deficiencies of our approximations are not accounted for.
The latter can be estimated by the differences between the results from
the two approximations used.

In order to compare these measured values to those predicted in the escape
model, we inject $N$ cosmic rays  at $z=0$ in the Galaxy and follow 
their trajectories $\vec x_i(t)$ until they reach the edge of the Galaxy.
As radial distribution of the injection points, we use 
\be \label{SNdist}
n(r) \propto \left( r/R_\odot \right)^{0.7}
\exp\left[-3.5(r-R_\odot)/R_\odot\right] 
\ee
with $[n]=$\,kpc$^{-2}$,
assuming that the surface density of CR sources follows the distribution
of supernova remnants in the Galaxy~\cite{Green:2013qta}. Here
$R_\odot=8.5$\,kpc is the distance of the Sun to the Galactic center.
We employ $n(z)=n_0\exp(-(z/z_{1/2})^2)$ as model for the gas distribution 
in the Galactic disk, where $z$ is the distance to the Galactic plane,
$n_0=0.3/$cm$^3$ at $R_\odot$  and $z_{1/2}= 0.21$\,kpc inspired by \cite{gas}. 
We set $n=10^{-4}$/cm$^3$ as minimum gas density up to the edge of 
the Milky Way at $|z|=10$\,kpc. 
Then we calculate the average grammage 
$\langle X\rangle =N^{-1}c\sum_{i=1}^N\int dt\,\rho(\vec x_i(t))$ 
summing up the density along the trajectories of individual CRs.
Since the grammage 
$X(E)\propto E^{-\delta}$ scales as the confinement time 
$\tau(E)\propto E^{-\delta}$,  this quantity serves also as an
indicator for changes in the CR intensity induced by a variation
of the CR leakage rate.

\subsection{Jansson-Farrar  model for the GMF}

Let us recall first how the properties of a turbulent magnetic field 
determine the propagation of charged particles in the diffusion picture,
before we discuss the specific case of the JF model. A turbulent magnetic
field is characterized by its power-spectrum $\mathcal{P}(\vec k)$.
The maximal length  $L_{\max}$ of the fluctuations and the correlation 
length $l_{\rm c}$ are connected by $l_{\rm c} = (\alpha-1) L_{\max}/ (2\alpha)$ for $\mathcal{P}(k)\propto k^{-\alpha}$.
Assuming that the turbulent field is isotropic, the slope of the  
power-spectrum 
$\mathcal{P}(k)\propto k^{-\alpha}$ determines the energy dependence of the
diffusion coefficient $D(E)$ in the limit $E\ll E_{\rm cr}$
as $D(E)\propto E^{2-\alpha}$ on distances $l\gg L_{\max}$. 
Here, the critical energy $E_{\rm cr}$ is defined by $R_{\rm L}(E_{\rm cr})=l_{\rm c}$
and thus the condition $E\ll E_{\rm cr}$ ensures large-angle scattering,
while the requirement $l\gg L_{\max}$ guarantees that features of anisotropic
diffusion are washed out.
Finally, we recall that the confinement time $\tau$ scales as the inverse of the 
diffusion coefficient.

In our previous work~\cite{PaperI}, we used the JF model for the 
regular and turbulent components of the Galactic magnetic field~\cite{J}, 
choosing as the maximal length of the fluctuations $L_{\max} = 10$\,pc.
Note that for Kolmogorov turbulence the maximal length of the fluctuations 
$L_{\max}$ and the correlation length $l_{\rm c}$ are connected by 
$L_{\max}=5l_{\rm c}$ and that the diffusion coefficient scales as 
$D(E)\propto E^{1/3}$  for $E\ll E_{\rm cr}$.
We considered two values of its root mean square (rms) strength, the 
original one suggested in~\cite{J} ($\beta=1$) and a second one rescaling 
it to one tenth of its original value ($\beta=1/10$). 

\begin{figure}
  \includegraphics[width=0.45\textwidth,angle=0]{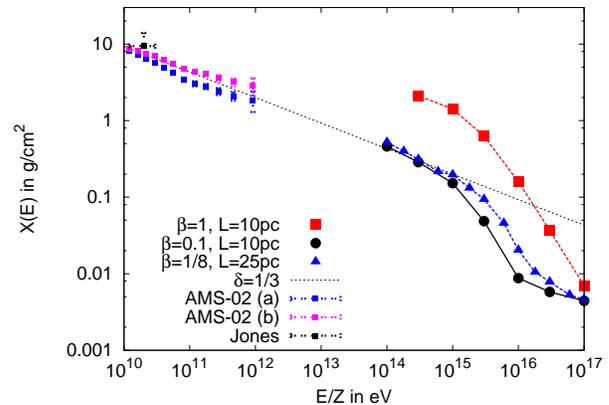}
  \caption{Grammage for different coherence lengths $l_{\rm c}$ and
    turbulent fields: red squares $L_{\max}=10$\,pc and $\beta=1$,
    black dots $L_{\max}=10$\,pc and $\beta=0.1$, and
    blue triangles $L_{\max}=25$\,pc and $\beta=0.125$; all cases for
   the JF GMF model~\cite{J}. Additionally we show
   the grammage deduced from B/C data.
\label{figX1}}
\end{figure}

In Fig.~\ref{figX1}, we compare the grammage calculated from simulated CR
trajectories for these two cases with the grammage deduced from B/C 
measurements.
Because of the large energy reach of the AMS-02 data, the extrapolation 
required from the lowest energy of our numerical calculations, 
$E=10^{14}$\,eV, to the measurements has decreased to two orders of magnitude. 
Using the JF model with $\beta=1$ as proposed in~\cite{J} would require a 
constant power spectrum of magnetic field fluctuations,  
$\mathcal{P}(k)\propto k^{-\alpha}$ with $\alpha=0$,
in the intermediate energy range. Such a power-spectrum is difficult to 
reconcile with the theoretical understanding of turbulence. 
Moreover, the CR spectrum is very close to a power-law above $\simeq 300$\,GV. 
This implies that if $D(E)$ would become significantly flatter beyond
TeV energies (e.g.\ changing from $D(E)\propto E^{1/3}$ to $\propto E^{0}$), 
then the injection spectrum of sources has to have the exact opposite 
change of slope (e.g.\ respectively from $\propto E^{-2.4}$ to 
$\propto E^{-2.7}$). Alternatively, a change in the source density
should compensate the change in $D(E)$ such that the observed CR 
intensity remains a nearly featureless power-law~\cite{Kachelriess:2005xh}. 
Although such a conspiracy cannot be excluded, it appears to us as a not 
very attractive option.

\begin{figure}
  \includegraphics[width=0.45\textwidth,angle=0]{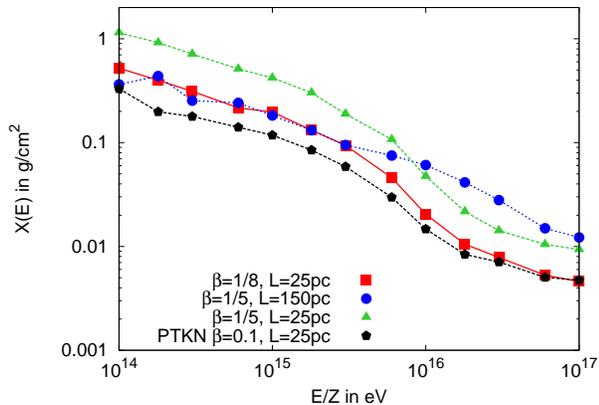}
  \caption{Grammage for different coherence lengths $l_{\rm c}$ and
    turbulent fields: 
     red squares $L_{\max}=25$\,pc and $\beta=1/8$,
     blue dots $L_{\max}=150$\,pc and $\beta=1/5$, and
     green triangles $L_{\max}=25$\,pc and $\beta=1/5$; all cases for
   the JF  model~\cite{J}. Additionally we show
   the grammage for the PTKN model with  $L_{\max}=25$\,pc and $\beta=0.1$
   by black stars.
\label{figX2}}
\end{figure}

Choosing a Kolmogorov\footnote{Note that 
a Kraichnan power-spectrum ($\alpha=3/2$)  would require a stronger
rescaling of the turbulent field, leading to a potential conflict with
synchrotron data.} 
power-spectrum $\mathcal{P}(k)\propto k^{-5/3}$ 
as the theoretical model with the smallest slope $\alpha$, we have to reduce 
$B_{\rm rms}$ therefore by a scaling factor $\beta<1$ relative to the
$B_{\rm rms}$ suggested in~\cite{J}. The exact value of 
$\beta$ depends on the chosen coherence length $l_{\rm c}$: A smaller
coherence length leads to faster diffusion and thus to a smaller value
of the grammage. For instance, a  coherence length close to the upper limits  
derived in Ref.~\cite{Iacobelli:2013fqa}, $l_{\rm c} = 5$\,pc allows
with $\beta=1/8$ a somewhat weaker reduction in the level of the 
turbulence,
cf.\ the blue line in Fig.~\ref{figX1}. Increasing the coherence 
length even further to $l_c=30$\,pc, the scaling factor can be reduced
to $\beta=1/5$, cf. Fig.~\ref{figX2}.

Next we examine how the shape of the grammage $X$ as function of energy 
$E/Z$ depends on the two parameters $l_c$ and $\beta$. In~\cite{PaperI}, 
we discovered a specific shape of $X(E)$ that leads not only to a
knee-like feature but reproduced also the recovery of the proton and 
helium spectra above $E/Z\sim 10^{16}$\,eV, visible in the KASCADE-Grande
data. From the examples in Figs.~\ref{figX1} and \ref{figX2}, it is 
clear that a too strong turbulent field, $\beta\sim 1$, results in
knee-like feature at too high energy. Compensating a relatively 
strong turbulent field by decreasing the coherence length tapers off
both the knee-like feature and the recovery, as shown by the case 
$l_c=30$\,pc in Fig.~\ref{figX2}. As a consequence, the allowed range of 
turbulent field strengths and coherence lengths is correlated and very 
restricted, $l_{\rm c} \simeq (1-10)$\,pc and $\beta  \simeq 1/10 - 1/8$.

\begin{figure}
  \includegraphics[width=0.35\textwidth,angle=270]{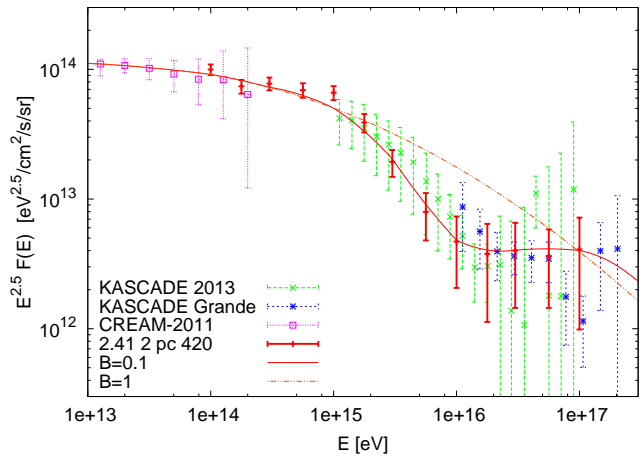}
  \caption{Intensity using $I(E)\propto X(E)$ (red solid line) for the reduced turbulent field compared to the case of full turbulent field (red dashed line). }
  \label{Fig_pFluxRecovery}
\end{figure}

This behavior is best illustrated comparing the modulation induced by
the energy dependence of $X(E)$ on the intensity of protons to KASCADE 
and  KASCADE-Grande data, and asking that a certain parameter choice 
reproduces the position and the shape of the proton knee.  
In Fig.~\ref{Fig_pFluxRecovery}, we show
$I(E)=I_0(E)X(E)$, where in $I_0(E)=I_0 (E/E_0)^{\alpha}$ the normalisation
$I_0$ and the slope $\alpha$ are chosen such to obtain a good agreement
with observations below the knee for the case of full ($\beta=1$, 
$L_{\max}=10$\,pc) and reduced ($\beta=0.1$) turbulent fields with 
$L_{\max}=10$\,pc:  
This comparison demonstrates that only the case 
with reduced turbulence can reproduce the observed shape of the 
proton flux.

\subsection{Pshirkov et al. model for the GMF}

In order to test the dependence of our results on the GMF model, we compute 
additionally the grammage $X(E)$ in the Pshirkov {\it et al.\/} 
model~\cite{P1,P2}.  Its regular field consists of toroidal components in the 
Galactic halo, and of a disk field which follows spiral arms in the 
Galactic plane. We choose the bisymmetric (BSS) benchmark model 
of~\cite{P1}, where the disk field presents reversals between consecutive 
arms. The authors of~\cite{P1,P2} did not present a three-dimensional model 
for the turbulent field. However, Pshirkov et al.~\cite{P2} derived upper 
bounds on the deflections of ultra-high energy CR (UHECR) induced by
the turbulence magnetic field. Requiring that these bounds are satisfied
allows us to  construct a toy model of the turbulent GMF.

We choose $l_{\rm c} = 5$\,pc and as profile function
$B_{\rm rms}(z) = B_{0} \exp(-|z|/1.8\,{\rm kpc})$.
Since the UHECR deflections at Earth as predicted by~\cite{P2} do not 
show any significant dependence on the Galactic longitude, we neglect
for simplicity a possible weak dependence of $B_{\rm rms}$ on the 
Galactocentric radius.

\begin{figure}
  \includegraphics[width=0.49\textwidth,angle=0]{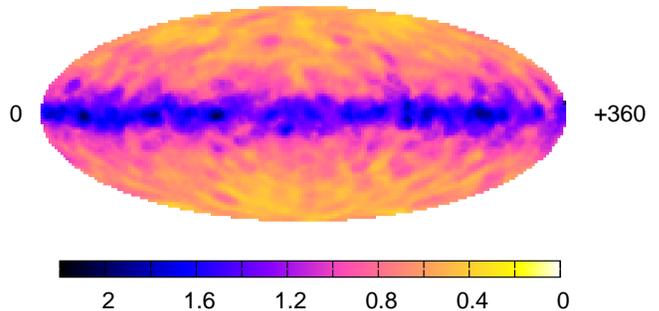}
  \caption{Deflections of 40\,EeV protons in a 
turbulent GMF realization with $L_{\max} = 100$\,pc and $B_{\rm rms} = 5\,\mu$G$\times \exp(-|z|/1.8\,{\rm kpc})$.}
  \label{Fig_UHECR_Defl}
\end{figure}

We backtrack individual 40\,EeV protons from the Earth in a realization 
of isotropic Kolmogorov turbulence with such characteristics\footnote{In order to be compatible with the asumptions of~\cite{P2}, we have used 
$l_{\rm c} = 50$\,pc for this comparison.}. We 
compute their deflections on the sky, and present them in 
Fig.~\ref{Fig_UHECR_Defl}, after smoothing over $5^\circ$ circles. 
One can see that 40\,EeV proton deflections in such a turbulence are 
compatible with the fit for the upper limit on deflections from~\cite{P2}, 
both at high and low Galactic latitudes. For such a profile, the constraints 
at high latitude are more stringent than those close to the Galactic plane. 
Therefore, the results of Ref.~\cite{P2} imply that, for a scale height of 1.8\,kpc of the turbulent field, $B_{0}$ should not be significantly larger than 
$\approx 5\,\mu$G. We take this value in the following.

We can now compute the grammage  in the PTKN model for the regular field,
which we supplement by our toy model for the turbulent field. We find that 
we have to reduce the normalisation $B_0$ of the turbulent field by a
similiar factor $\beta$ as in the JF model: The case $\beta=1/10$
and $l_{\rm c}=5$\,pc is shown in  Fig.~\ref{figX2} with black squares. Compared
to our favorite cases in the JF model ($\{\beta=1/8,l_{\rm c}=5{\rm pc}\}$
and $\{\beta=1/10,l_{\rm c}=2{\rm pc}\}$), the qualitative behavior of
$X(E)$ is very similar. It is therefore possible to limit the numbers
of models, and we then decide to use the JF model for the rest of this study.

\section{Fluxes of Galactic CR groups}
\label{IndividualFluxes}

\subsection{Diffuse fluxes from all sources}

\begin{figure*}
  \centerline{\includegraphics[width=0.35\textwidth,angle=270]{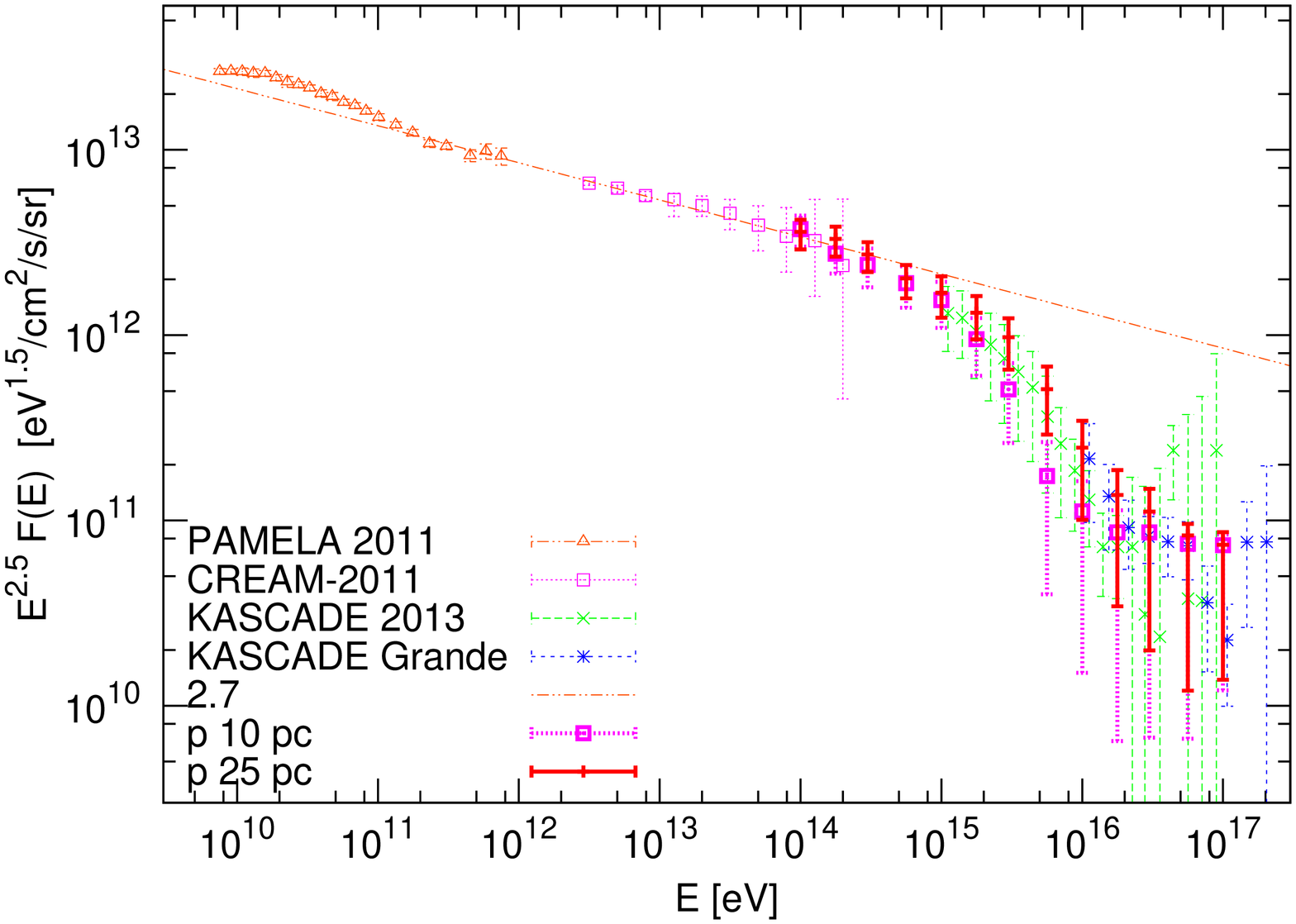}
              \hfil
              \includegraphics[width=0.35\textwidth,angle=270]{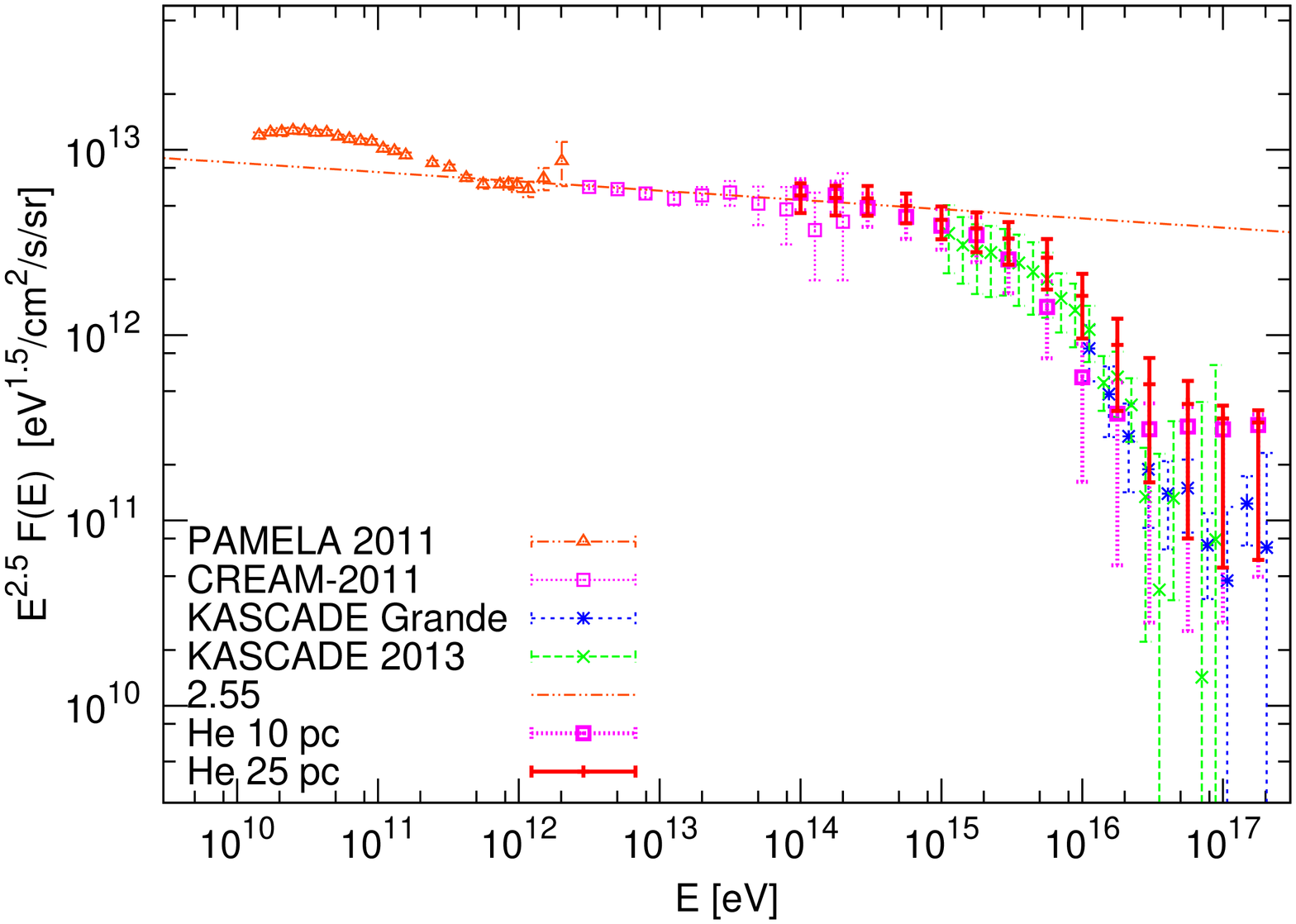}
             }
  \centerline{
              \includegraphics[width=0.35\textwidth,angle=270]{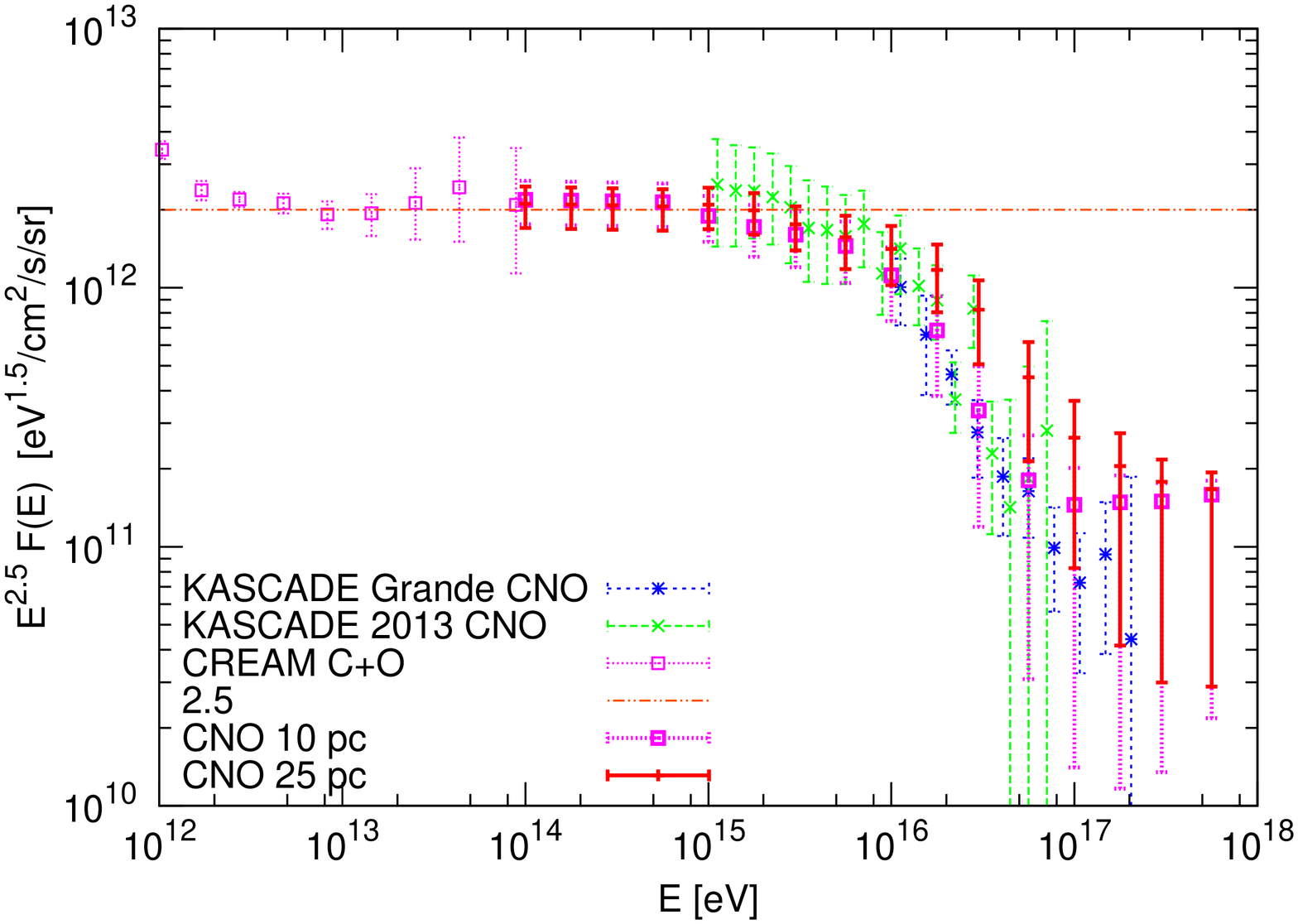}
              \hfil
              \includegraphics[width=0.35\textwidth,angle=270]{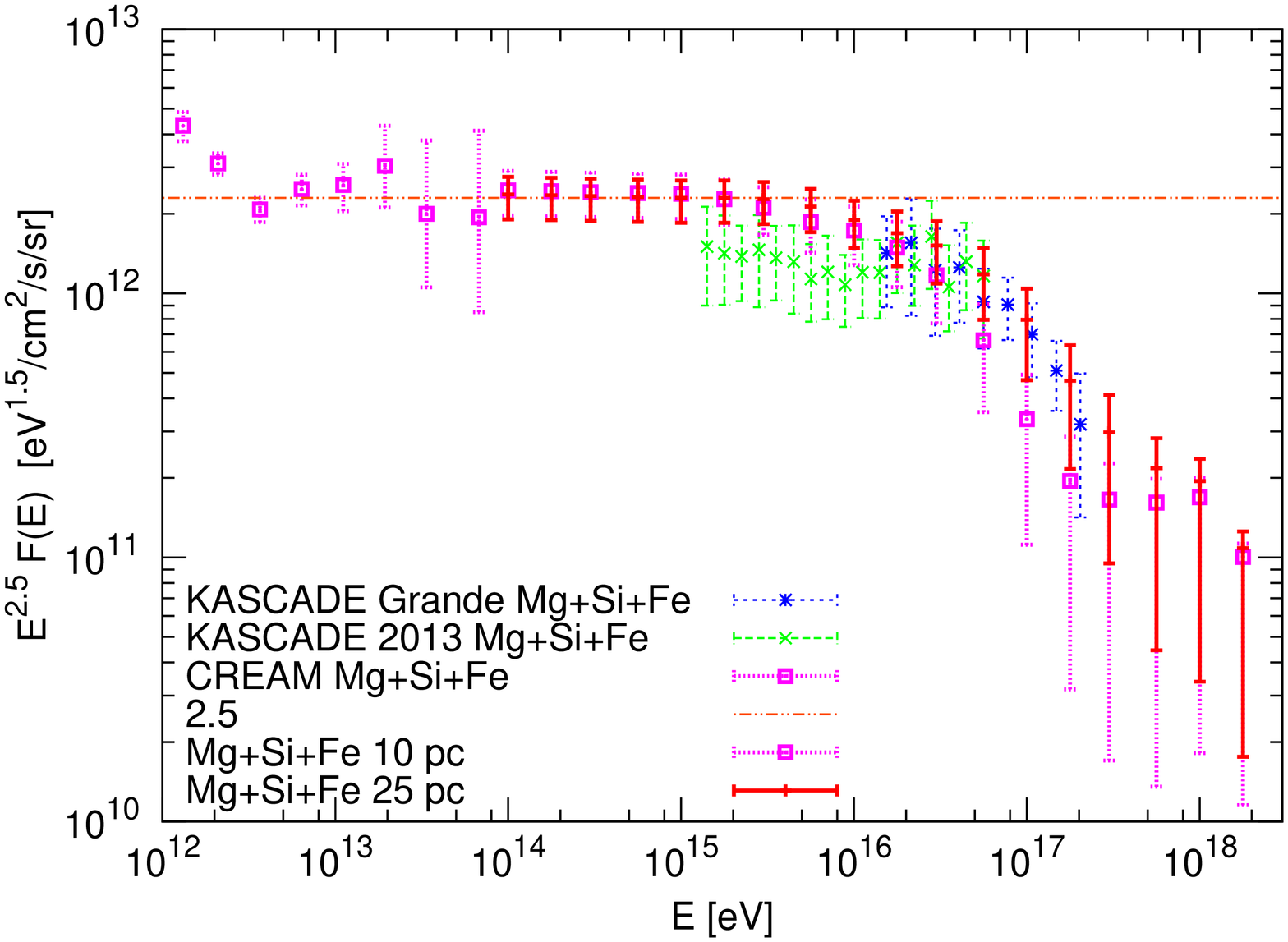}
             }
  \caption{Fluxes of CR protons (upper left), helium (upper right), CNO (lower left) and heavy elements including Mg, Si and Fe (lower right) are shown in red. Errorbars show the variations in time of the fluxes. Experimental data from PAMELA for protons and helium, CREAM~\cite{CREAM}, KASCADE~\cite{dataKG} and KASCADE-Grande~\cite{dataKG}.}
\label{Fig_4Fluxes}
\end{figure*}

In order to calculate the CR flux at Earth, we adopt the following procedure. First, we compute the three-dimensional CR density around a source at different times, and later use it as a template to infer the CR distributions around other sources. This allows us to reduce the required computing time and makes the problem tractable. For this template, we take a source located in the vicinity of the Earth and propagate individual CR protons from it in the GMF models which are presented in the previous Section. We do this for CRs with energies between 100\,TeV and 100\,PeV, and take four energies per decade. 
We divide the space around the source in cylindrical sectors with radii 
ranging from 0.1\,kpc to 4\,kpc. We checked that the remaining contribution from the radial range $4-10$\,kpc does not 
change our results. Their widths in vertical direction depend 
on $z$ and we choose their boundary lines as $|z|=100$\,pc, 200\,pc, 300\,pc, 
400\,pc, 500\,pc, 700\,pc, 1\,kpc, 2\,kpc, 3\,kpc, 5\,kpc, and 10\,kpc. 
We divide time in bins of 5\,kyr with a total of 6000 bins, up to $t=30$\,Myr. 
We checked also that the chosen bin sizes are small enough, in order to have
no impact on the results. 
For every propagated CR, we save the fraction of its path in a given bin and average over all simulated particles. From this, we deduce the three-dimensional time-dependent CR density in the Galaxy for this source, and then for a distribution of sources.

We create the ensemble of CR sources as follows. 
We generate their positions within the Galactic disk ($|z|<100$pc), assuming that the density of CR sources follows the distribution given by Eq.~(\ref{SNdist}). This distribution is assumed to depend only on the distance to the Galactic center, and not on the direction around it in the plane. The remaining parameters are the frequency of CR sources and the energy released in CRs by each of them. Only the product of these parameters is constrained by observations. We fix the energy released in CRs by each source to $E_{\rm tot}=10^{50}$\,erg, leaving the source frequency as the only free parameter. We sum up the contributions from these generated sources to the CR flux at Earth, in any time bin, and for a total duration of 300\,Myr. 
For each energy bin, we save both the average flux at Earth and its one sigma 
deviation in time, i.e.\ the values of the flux where 16\% of the cases are above and below the average. Note, that the upper and lower limits are not symmetric at high energies, since there are more cases with lower flux than average.

Results for nuclei with charge $Z$ are deduced from the above calculations 
for protons by shifting the energy by a factor $Z$. We then interpolate the 
resulting CR nuclei fluxes to the same energies as for protons. 
At energies below $Z \times 100$\,TeV, we assume that diffusion in the 
Kolmogorov turbulence shifts the injection power-law $\propto E^{-\alpha}$
by 1/3 to the spectrum $\propto E^{-\alpha-1/3}$ observed. This is indeed 
what we observe in our simulations in the energy range $\simeq Z \times (100-300)$\,TeV. Therefore we assume that the CR spectrum of protons released by 
sources follows a power-law spectra $\propto E^{-2.4}$; the maximal energy 
$E_p$ of protons will be fixed later by considering constraints form the 
resulting dipole anisotropy and the observed nuclear composition of CRs.
For all other nuclei, we use power-law spectra with either $\propto E^{-2.17}$ 
or $\propto E^{-2.22}$ and maximal energy $ZE_p$. These power-law indices are chosen so as to fit the direct observations from CREAM at low energy. We fix 
the density of sources by normalizing the flux found in our simulations to 
the observed 
one at 100\,TeV. On average, we require 440\,sources per 100\,kyr for a 
total energy per source of $E_{\rm tot}=10^{50}$\,erg, so as to fit the 
observed CR spectra. Within one time bin of 5 kyr we generate sources according to a Poisson distribution with an average of 22 sources, corresponding to the required source density on larger time scales.

In Fig.~\ref{Fig_4Fluxes}, we plot the CR nuclei fluxes, multiplied by $E^{2.5}$, as a function of energy. In the upper left and upper right panels of Fig.~\ref{Fig_4Fluxes}, we show the proton and helium fluxes, both for turbulent fields with $L_{\max} = 10$\,pc and with $L_{\max} = 25$\,pc. We plot orange lines $\propto E^{-2.4-1/3}$ (upper left panel) and $\propto E^{-2.22-1/3}$ (upper right panel), which represent the slopes expected theoretically at Earth, for our injection spectra with $\alpha=2.4$ and $2.22$ as power indices and Kolmogorov 
turbulence. Note that the slopes of the injection spectra required for nuclei, 
$\alpha\simeq 2.2$, coincide with the naive expectations from diffusive shock
acceleration. Only the proton injection spectra requires a somewhat softer
slope, $\alpha=2.4$, than expected.

In the two upper panels, we show the experimental data from PAMELA~\cite{PAMELA} (orange points), CREAM~\cite{CREAM} (magenta), KASCADE~\cite{dataKG} (green) and KASCADE-Grande~\cite{dataKG} (blue). The proton flux reported by KASCADE-Grande is 40\% larger than the flux from KASCADE in the $(10-30)$\,PeV region, where the errorbars of both experiments are relatively small. In contrast, the helium flux from KASCADE-Grande is below the one measured by KASCADE . This behavior may be explained by the insufficient discrimination power between protons and helium in the KASCADE-Grande experiment~\cite{AH}. For this study, we choose therefore to reduce the proton flux of KASCADE-Grande by 40\%, and add this difference to the helium flux, in same energy bins. By doing so, the CR fluxes of KASCADE-Grande and KASCADE experiments become consistent with each other.

In the lower left panel of Fig.~\ref{Fig_4Fluxes}, we plot the CNO flux, which predominantly consists of carbon and oxygen. We calculate the carbon and oxygen fluxes by normalizing them to the CREAM fluxes interpolated to higher energies with power-laws, and then sum them up. The CREAM flux in this figure is the sum of its carbon and oxygen fluxes, where we use carbon energy bins for the binning, and interpolate the oxygen flux to these bins before summing up. KASCADE and KASCADE-Grande measurements of the CNO flux are directly compared to our fluxes.

\begin{figure*}
  \centerline{
              \includegraphics[width=0.35\textwidth,angle=270]{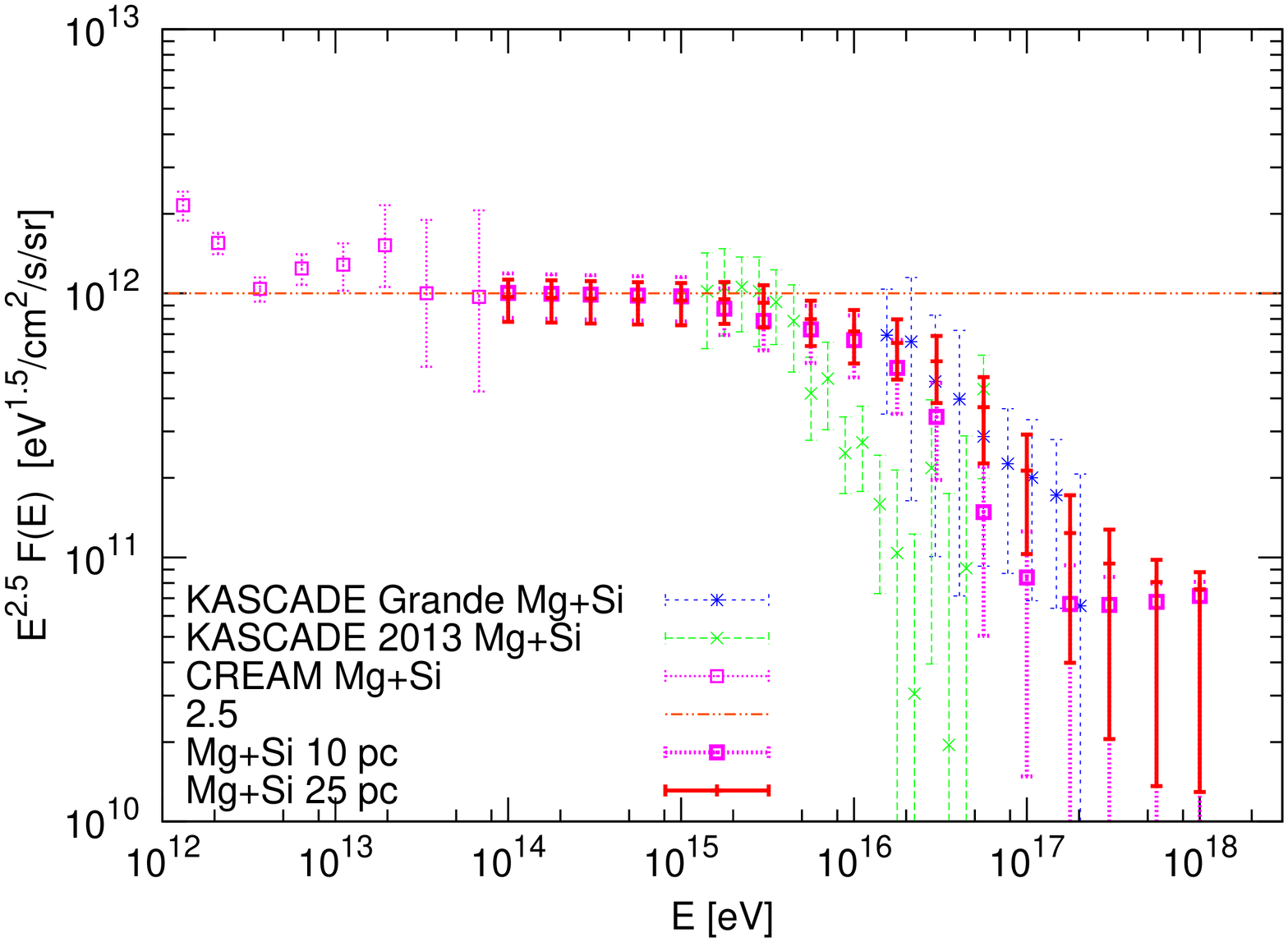}
              \hfil
              \includegraphics[width=0.35\textwidth,angle=270]{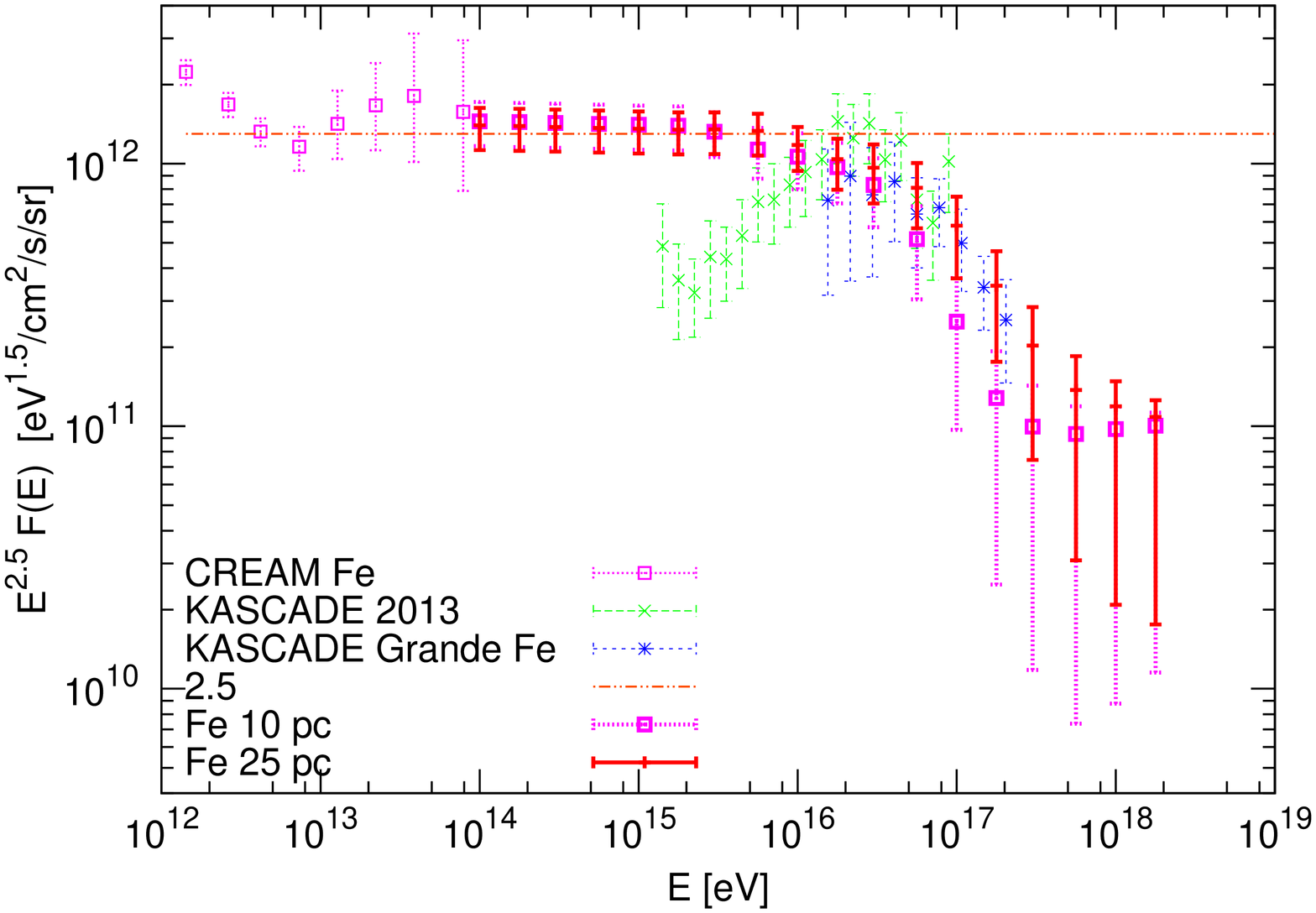}
             }
  \caption{Fluxes of Mg and Si (left) and Fe (right) are shown in red for $l_{\rm max}=25$\,pc and in blue for $l_{\rm max}=10$\,pc. Errorbars show the variations in time. Experimental data from CREAM, KASCADE and KASCADE-Grande.}
\label{Fig_FluxesMgSiFe}
\end{figure*}

In the lower right panel of Fig.~\ref{Fig_4Fluxes}, we show the flux of heavy nuclei, which is dominated by Mg, Si and Fe nuclei. Since the KASCADE and KASCADE-Grande collaborations divide this flux in two parts (\{Mg + Si\} and Fe), we also plot these two contributions separately in Fig.~\ref{Fig_FluxesMgSiFe}: Mg and Si in the left panel, and Fe in the right panel. It is possible to link the last points from CREAM to the first points from KASCADE-Grande with a smooth power-law, but not to those from KASCADE. This is likely to be due to the difficulty for KASCADE to distinguish between Si and Fe nuclei~\cite{AH}. Therefore we choose to sum up the Mg, Si and Fe fluxes into a single 'heavy component' in the lower right panel of Fig.~\ref{Fig_4Fluxes}. In this figure, the combined 'heavy nuclei flux' as measured by the KASCADE experiment is smooth and agrees well with a simple power-law extrapolation of the CREAM flux to higher energies. It also agrees with the KASCADE-Grande flux. As for the other components, the model presented in this work fits well the heavy nuclei flux too.

As can be seen in Fig.~\ref{Fig_4Fluxes} (upper left), the CR proton flux follows a power-law from 300\,GeV up to about 1\,PeV. It then changes to a steeper slope at the knee, and recovers at $\simeq 10$\,PeV to a flatter power-law with index $\alpha \simeq 2.5$. Similar 'knee-like' cutoffs, shifted by factors $Z$ in energy, are visible in the fluxes of all groups of CR nuclei---see the other panels of Fig.~\ref{Fig_4Fluxes}. These plots demonstrate that the ``escape model'' fits very well all these observations. As discussed previously in~\cite{PaperI}, the knee is due, in this model, to a change in behaviour with energy of the CR diffusion coefficient. The energy of the knee corresponds to the energy at which the Larmor radius of CR protons is of the order of the coherence length of the turbulent magnetic field ($l_{\rm c}=L_{\max}/5$ for a Kolmogorov spectrum). For the field strengths we consider in this paper, $B_{\rm rms} \simeq 0.3\,\mu$G (or $\beta=1/8$) averaged over a circle of radius 1\,kpc distance from the solar position in the Galactic plane, we find in our calculations a change in the slope of the CR flux at about 1\,PeV, as observed in the {\em proton\/} data.

\subsection{Flux from nearby sources}

\begin{figure*}
  \includegraphics[width=0.34\textwidth,angle=270]{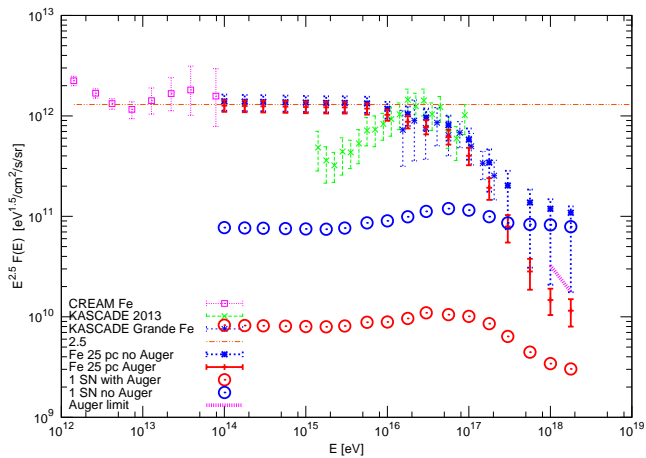}
  \includegraphics[width=0.34\textwidth,angle=270]{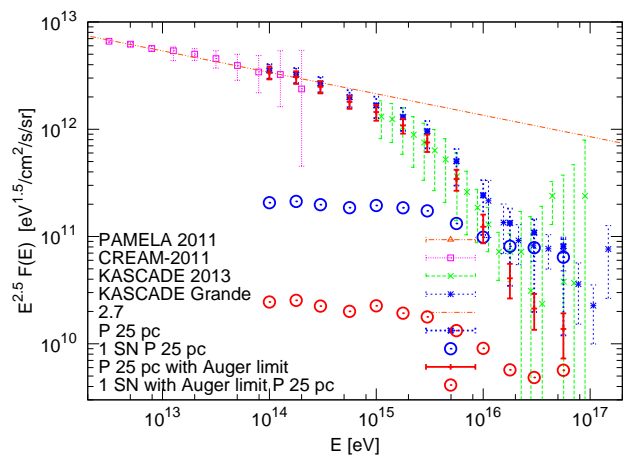}
  \caption{Left panel: Experimental data for iron together with the 
   predicted iron flux without (blue errorbars) and with the
   Auger constraint (red errorbars). Additionally, 
   the iron flux from the dominant source is shown without
   (blue circles) and  with Auger constraint (red circles). 
   Right panel: The same plot for protons.
   All cases for  $l_{\rm max}=25$\,pc.
  \label{Fig_1s}}
\end{figure*}

In the escape model, the flatter part of the CR proton flux above $\simeq 10$\,PeV is dominated by recent nearby sources. This is due to the fact that the confinement time of CR protons in the Galaxy quickly drops with energy beyond the energy of the knee. 
The  high energy  part, $E\gsim3 \times 10^{17}$\,eV,  of the Galactic flux is 
dominated by heavy elements (Mg+Si+Fe).  Recent nearby sources would dominate 
the flux of heavy elements at these energies. The composition study 
published  by Auger in~\cite{Aab:2014aea} constrains however the fraction 
of iron.
Using conservatively the results obtained using the EPOS-LHC simulation, 
the iron fraction above $6\times 10^{17}$\,eV is limited as $\lsim 20\%$.
We can add this constraint,
excluding all time bins where the iron fraction exceeds this Auger limit.
In the left panel of Fig.~\ref{Fig_1s}, we show  experimental data 
for iron from CREAM, KASCADE and KASCADE-Grande together with the predicted 
iron flux without (blue errorbars) 
and with accounting of the Auger iron constraint (red errorbars). 
The maximally allowed iron flux using the composition constraint 
from~\cite{Aab:2014aea} is show as a magenta line.
Since the signature of nearby, recent CR sources is a large iron fraction,
the Auger constraint effectively eliminates these cases, resulting in
a reduced flux at high energies, $E\gsim 10^{17}$\,eV.
In the right panel of Fig.~\ref{Fig_1s}, we show the same plot for
protons.

Next we consider the effect of the  Auger limit on the flux of the source
which gives the maximal contribution at highest energy 100\,PeV.
The flux from this dominant source is shown in both panels of Fig.~\ref{Fig_1s}
without (blue circles) and  
with (red circles) accounting for Auger constraint on the iron fraction.
In the former case, the dominant source contributes almost 100\% of the
proton flux at $5\times 10^{16}$\,eV, while taking the Auger iron constraint
into account reduces the dominance of the strongest source. Clearly,
the relatively small fraction of iron observed by Auger disfavors
the presence of a dominating source even at the end of Galactic CR
spectrum.

Finally, we note that in the cases when the Auger iron constraint is violated,
the total proton flux exceeds then the measured one and the 
knee-like structure is less pronounced than in the observed data.
Above $Z\times 10^{16}$\,eV,  the predicted CR flux is dominated by a
single nearby and recent source. Such a situation contradicts not only 
the Auger iron limit, but would violate also the limits on the dipole 
anistropy of the CR flux:
If one assumes that Galactic sources are able to accelerate only to 
energy just below the Auger constraint on iron, 
$E_{\rm max}<7\times 10^{17}$\,eV, the constribution of recent nearby sources
is still limited by the Auger anisotropy limits, cf.\ the discussion in the
next section. Thus the Auger limits 
on anisotropy and the iron fraction exclude the possibility that 
the highest energy part of the Galactic CR spectrum is dominated by a
recent nearby source.

\section{Transition from Galactic to extragalactic CRs}
\label{Transition}

\begin{figure}
  \includegraphics[width=0.35\textwidth,angle=270]{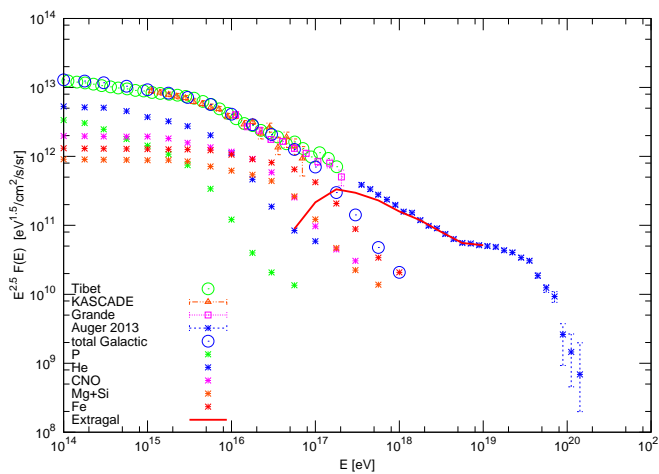}
  \caption{CR flux and the extragalactic component we predict 
           for $\Rm=1\times 10^{17}$\,eV, shown with the data
           from 
CREAM~\cite{CREAM}, 
KASCADE~\cite{Antoni:2005wq},
KASCADE-Grande~\cite{Apel:2012rm},
TIBET~\cite{Amenomori:2008aa}, and
Auger~\cite{Aab:2013ika} 
experiments.}
  \label{Fig_AllPartExtragal}
\end{figure}

\begin{figure}
  \includegraphics[width=0.47\textwidth,angle=0]{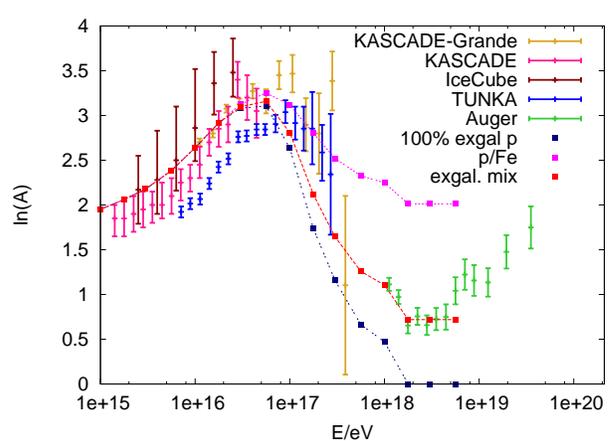}
  \caption{Average of the logarithmic mass ln\,A predicted by our model
  for three different assumptions on the composition of extragalactic CRs, 
   versus the experimental data.}
  \label{Fig_lnA}
\end{figure}


Determining at which energy $E_\ast$ the CR flux starts to be dominated by
extragalactic sources is one of the most important unsolved  problems in CR 
physics. While in the 'dip model' of Ref.~\cite{dip} the transition energy 
is as low as $E_\ast \ap {\rm a~few} \times 10^{17}$\,eV, the ankle has been
in other models identified with the transition between Galactic and
extragalactic CRs~\cite{ankle}, 
$E_\ast \ap E_a\ap {\rm a~few} \times 10^{18}$\,eV. In our model, 
the energy of the transition $E_\ast$ depends both on the maximum rigidity 
$\Rm$ to which Galactic  sources are able to accelerate CRs 
and on the distance to the nearest active source. As we have seen in
the previous section, the Auger constraint on the iron fraction
effectively eliminates the possibility that a single source dominates
the high-energy part of the Galactic CR spectrum, reducing thereby
the fluctuations.

We impose in the following the iron constraint throughout and assume 
that the end of the Galactic CR spectrum is determined by 
the maximum rigidity $\Rm$ to which Galactic sources are able to 
accelerate CRs. Although  $\Rm$ is a free parameter in our model, 
the energy of the transition $E_\ast$  can be determined by using 
additional observational constraints. One possibility is to constrain 
the maximum contribution of Galactic 
sources to the total CR flux by using the observational limits on the 
anisotropy of the CR flux. Another way to determine $E_\ast$, 
is to study the elemental composition of primary CRs and to use the fact 
that the composition 
of Galactic and extragalactic CRs should in principle differ from 
one another.

We start with the latter method. As a first step,  we derive the all-particle
CR flux summing up all CR groups and compare it to the experimental data of
KASCADE~\cite{Antoni:2005wq}, KASCADE-Grande~\cite{Apel:2012rm},
TIBET~\cite{Amenomori:2008aa}, and
Auger~\cite{Aab:2013ika}. 
Then we deduce the extragalactic flux for a given $\Rm$ by subtracting 
the predicted total Galactic flux from the measured total CR flux.
The resulting extragalactic flux is shown in Fig.~\ref{Fig_AllPartExtragal} 
with a red solid line for $\Rm=1.0\times 10^{17}$\,V. Next, 
we have to fix the nuclear composition of the extragalactic CR
flux. As a first approximation, we can assume that its composition is
constant in a sufficient small energy interval around $E_\ast$. In contrast, 
the Galactic CR composition is strongly energy dependent between the
knee and the cutoff of the Galactic flux. Thus, we expect
that an observable like the average of the logarithmic mass number, 
$\ln(A)$, will be quickly changing for $E\lsim E_\ast$, while being 
approximately constant for energies slightly above $E_\ast$.

In Fig.~\ref{Fig_lnA}, we plot measurements of $\ln(A)$ from several 
experiments, together with the values of $\ln(A)$ calculated within the 
'escape model' studied here.  The points for KASCADE have been computed 
by converting the flux measurements given in~\cite{dataKG} into 
$\ln(A)$ values\footnote{Recall that the KASCADE data for the heavy components
showed a discrepancy to the extrapolation of the CREAM and 
KASCADE-Grande data and we had to sum them into a single heavy component.
For the calculation of $\ln(A)$, we used instead the original fluxes for 
the separate CR groups what explains the small off-set between our prediction
and $\ln(A)$ deduced from KASCADE data at low energies.}.
The most striking feature, namely the peak in $\ln(A)$ around 
$5\times 10^{16}$\,eV, is clearly visible in all data sets, although its 
exact position and strength depend on the experiment. Our model reproduces 
the trend in the data very well. At higher energies, 
the composition becomes lighter because of the 'flattening' of the escape 
time at such energies, see Section~\ref{GMF}. For the value of $\Rm$ we consider here, 
extragalactic CRs start to contribute to the observed flux at 
$\approx 10^{17}$\,eV. Consequently, above this energy, the exact value and shape of $\ln(A)$ 
depends on the assumed composition of the extragalactic flux. 
In blue, we show $\ln(A)$ for an extragalactic flux made of protons only, 
in magenta for a mix of 50\% p and 50\% Fe, and in red for a mix of 60\% p, 25\% He, and 15\% N. 
Independently of the composition chosen for the extragalactic component, we can identify the energy 
where $\ln(A)$ stops to decrease with the maximum energy $E_{\rm max,Fe}$ to 
which Galactic sources can accelerate iron. It corresponds to the rigidity $\Rm=E_{\rm max,Fe}/26e$. 
This transition is clearly visible in the PAO data, around the ankle, and allows us to 
determine the maximum rigidity as $\Rm\simeq 1\times 10^{17}$\,V. 

Accelerating Galactic CRs to $\Rm\simeq 1\times 10^{17}$\,V is challenging, even for supernovae 
exploding in dense winds. However, several models in which one
may reach such a high energy have been proposed. One possibility 
is the two-step acceleration of CRs in OB regions, see 
Ref.~\cite{Parizot:2014ixa} for a recent review. Since most of 
core-collapse SNe are located in superbubbles, CRs accelerated  
by individual SN remnants may be additionally accelerated in superbubbles 
to energies $\Rm\simeq 1\times 10^{17}$\,V~\cite{Parizot:2014ixa}.
As another possibility, Ref.~\cite{Lemoine:2014ala} suggests
that CRs can be accelerated to ultra-high energies at the termination 
shock of young pulsar winds. Note that in the  early stages when 
the acceleration is most effective pulsars stay in the same OB regions, 
and the argument discussed applies in this case as well.
The TeV gamma-ray emission from extended Galactic sources was studied 
in Ref.~\cite{Neronov:2012kz}. There it was found that the number of 
extended  sources detected in Fermi data is consistent with the expected 
number of TeV CR sources. The majority of these TeV gamma-ray sources 
was associated with pulsars. If these gamma-rays have a hadronic origin,
pulsars may be candidates for the Galactic CR sources.

For the case of a mixture of 60\% p, 25\% He, and 15\% N (red curve in Fig.~\ref{Fig_lnA}), 
we obtain a good agreement with the $\ln(A)$ data from PAO up 
to $2\times 10^{18}$\,eV. While this choice of composition is not unique, it is 
consistent with the results from the recent composition study published in~\cite{Aab:2014aea}. 
In particular, Ref.~\cite{Aab:2014aea} found the fraction of iron to be below 20\% 
above $6\times 10^{17}$\,eV. This agrees well with the results of our model, 
where the Galactic flux at $6\times 10^{17}$\,eV consists purely of iron 
but contributes to only 15\% of the total CR flux.

\begin{figure}
  \includegraphics[width=0.47\textwidth,angle=0]{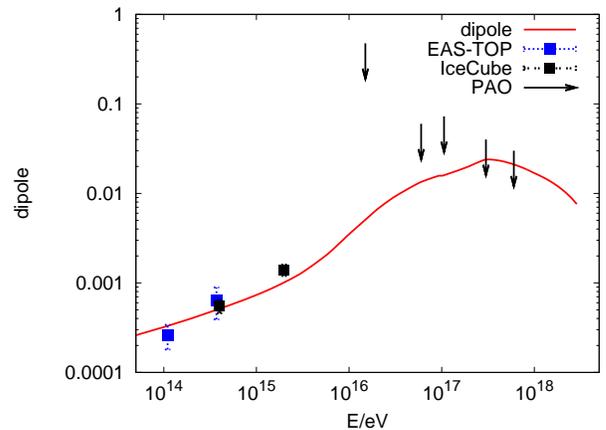}
  \caption{Dipole amplitude $d(E)$ as a function of energy 
    $E$ in the GMF model of Ref.~\cite{J}, using a 
    reduced turbulent magnetic field with $\beta=1/8$ and $L_{\max}=25$\,pc.}
  \label{figd}
\end{figure}


In addition to fitting the above observables, we still have to verify that the 
model presented here is also consistent with the existing upper limits on the CR anisotropy.
In the diffusion approximation, the CR dipole anisotropy $d$ is given by 
$\vec d = 3D \: \vec\nabla\ln(n) /c$. Following the same procedure as in
\cite{PaperI}, we compute the average anisotropy and derive 
the energy dependence of $D(E)$ from the escape rate as calculated 
previously, setting $D(E/Z)\propto1/\tau_{\rm esc}(E/Z)$. We fix the 
proportionality constant by requiring that the dipole amplitude 
$d=\sum_k f_k d_k$ equals the dipole component $\tilde d$ observed by 
the EAS-TOP collaboration at $E=1.1\times 10^{14}$\,eV~\cite{EASTOPD,EP}. 
Here, $k$ labels the groups of nuclei we consider in the Galactic
flux plus an extragalactic component. The latter has a dipole amplitude
which is independent of its composition and which we set equal to 
0.6\%, as expected for the extragalactic Compton-Getting 
effect~\cite{Kachelriess:2006aq}. The factor $f_k$ corresponds to
the fraction the component $k$ contributes to the total CR flux, 
and $d_k\propto 1/\tau_{\rm esc}(E/Z)$ to their individual dipole. The relatively low value of the CR dipole 
measurements at TeV--PeV energies is known as the 'CR anisotropy problem'. Some authors have suggested that 
conditions of the local interstellar turbulence may be the cause~\cite{Zirakashvili2005,Mertsch:2014cua}.

In Fig.~\ref{figd}, we show the resulting dipole amplitude $d$ as a 
function of energy $E$. As expected, the amplitude rises below the 
knee as $E^{1/3}$, while it increases approximately as $E^{0.7}$ until
$1\times 10^{17}$\,eV. At higher energies, the dipole amplitude decreases, which is due to the facts that 
the Galactic composition becomes heavier and that the
extragalactic contribution grows. We also plot the values of $\tilde d$ 
observed by IceCube~\cite{Ianisp}, as well as the 99\% C.L. upper limits 
on $d_\perp$ from the Pierre Auger Observatory~\cite{PAOaniso}. 
Comparing our estimate for the dipole amplitude with the upper limits 
in the energy range $10^{17}-10^{18}$\,eV, we should take into account that
the approximation $d\propto 1/\tau_{\rm esc}(E/Z)$ starts to break down above 
$E/Z \gtrsim 10^{17}$\,eV, which leads to a sizeable error. We conclude therefore that our 
prediction is marginally consistent with these limits. The Pierre 
Auger Observatory should however be able to reach a detection of the dipole anisotropy. 
Let us also note that the escape model predicts that the phase of the dipole 
amplitude varies strongly in the energy range between $1\times 10^{17}$ and $3\times 10^{18}$\,eV: This 
corresponds to the range where the transition from Galactic to extragalactic CRs lies. 
Such a picture is supported by current observations of the phase of the dipole, see Refs.~\cite{EASTOPD,Ianisp,PAOaniso}.

In summary, there are two reasons for having an early transition, from predominantly 
Galactic to predominantly extragalactic CRs, at $E \approx {\rm  a~few} \times 10^{17}$\,eV. First, the limits
on the observed dipole anisotropy requires either a very heavy Galactic
composition or a predominantly extragalactic contribution at $E \gsim 10^{18}$\,eV~\cite{Giacinti:2011ww,PAOaniso}. The former possibility is however strongly
disfavored by the recent composition measurements from the Auger 
collaboration~\cite{Aab:2014kda,Aab:2014aea}. Second, identifying the energy
where $\ln(A)$ stops decreasing with the maximum energy to
which Galactic sources can accelerate iron,
$E_{\rm max,Fe}\ap 3\times 10^{18}$\,eV, suggests that 
the maximal rigidity reached in Galactic sources satisfies 
$\Rm = E_{\rm max,Fe}/(26e) \sim 10^{17}$\,V.

\begin{figure}
  \includegraphics[width=0.35\textwidth,angle=270]{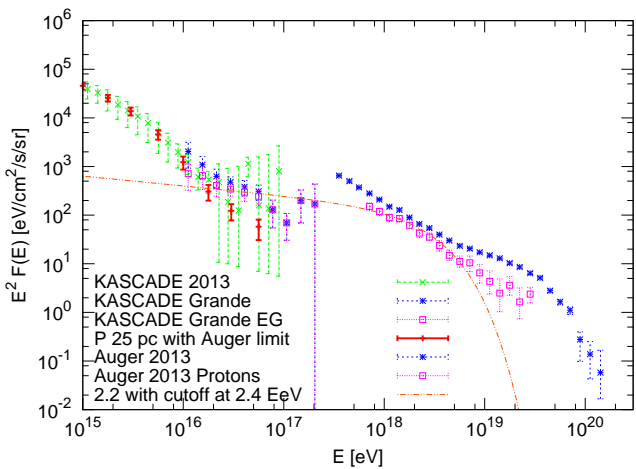}
  \caption{The extragalactic proton flux deduced
   from the Auger and KASCADE-Grande data is shown in magenta together
   with the proton flux observed by KASCADE (green errorbars) and 
  KASCADE-Grande (blue errorbars). The total CR flux from Auger
   (blue errorbars) is shown together with the deduced proton flux 
  (magenta errorbars).
  \label{Fig_exprot} }
\end{figure}

Finally, we comment on the contribution of extragalactic
protons to the observed proton flux by KASCADE and KASCADE-Grande.  
In Fig.~\ref{Fig_exprot}, we show these experimental data together
with the predicted Galactic proton flux (red errorbars), taking into account the 
Auger iron constraint. At $E\gsim 3\times 10^{16}$\,eV, the predicted Galactic 
proton flux lies below the measured one: Within the escape 
model, this difference should be accounted for by extragalactic protons.
Subtracting the measured proton flux from the flux calculated
in the escape model, we obtain a prediction for the extragalactic
proton flux shown in magenta in Fig.~\ref{Fig_exprot}.
Note that the absolute value of this extragalactic proton flux is
too small to impact the $\ln(A)$ plot, Fig.~\ref{Fig_lnA}. We can
check if the interpretation of this Galactic proton deficit in
our model as an extragalactic flux makes sense comparing it to
expectations at higher energies. We show therefore in Fig.~\ref{Fig_exprot}
additionally the total CR flux (blue) measured by Auger.
Applying the proton fraction from Ref.~\cite{Aab:2014aea} obtained 
using the EPOS-LHC simulation we can then derive the resulting proton 
flux which is shown in magenta. To guide the eye, we plotted also
a $E^{-2.2}$ power-law with an exponential cutoff at $E=2.4\times 10^{18}$\,eV
as an orange line.  Such a $E^{-2.2}$ power-law interpolates nicely between
our prediction for the extragalactic proton flux using the KASCADE-Grande
data and using the Auger data. This conclusion would not change using the 
spectrum and the composition  measured by the Telescope Array (TA), since
the two experiments agree  on the points that are the 
most important for our analysis: A very small fraction of iron 
and a large fraction of protons below $10^{19}$\,eV.
We conclude therefore that the extragalactic proton flux
determined in the escape model, although with large errors, is consistent
with the slope expected from shock acceleration and fits to the proton
flux determined by Auger below the ankle.

\section{Discussion}
\label{Discussion}

Before we conclude, we review the main properties of the proposed
escape model for Galactic CRs and the resulting consequences for
the transition between Galactic and extragalactic CRs.

\subsection{Constraints on the GMF}

The ``escape model'' which aims at explaining the CR data from 
$E/Z\sim 300$\,GeV to 100\,PeV by the energy-dependent CR leakage 
from the Milky Way is strongly constrained by experimental data:
\begin{enumerate}
\item
The position of the knee, $E_{\rm k} \approx 4$\,PeV , fixes a combination
of the coherence length $l_{\rm c}$ and the strength of the magnetic field. 
Approximately, these parameters have to satisfy 
$l_{\rm c} \sim R_{\rm L}(E_{\rm k})$, while
our numerical calculations show that the coherence length should lie 
in the range $l_{\rm c} = (2-5)$\,pc for acceptable magnetic field strengths.
\item
The shape of the knee and, for large $\Rm$, the subsequent recovery observed 
in the energy spectra of individual CR groups measured
by KASCADE-Grande can be reproduced only, if the turbulent magnetic field 
strength is smaller than assumed e.g.\ in the JF model of Ref.~\cite{J}, 
cf.~Fig.~\ref{Fig_pFluxRecovery}.
\item
A smaller strength of the turbulent field is also supported by  B/C data: 
They constrain the grammage 
traversed by CRs and are consistent with a Kolmogorov-like power-spectrum 
$\mathcal{P}(k)\propto k^{-5/3}$ of the turbulent field modes. Consistency 
with these measurements at lower energies also forces us to decrease 
$B_{\rm rms}$. More quantitatively, we have to reduce the turbulent field in 
the JF model  by a factor $\sim 8$, keeping the regular field unchanged.
\item
Such a reduction is in line with our determination of the diffusion 
coefficient in a purely turbulent magnetic field with strength 
$B_{\rm rms}=4\,\mu$G \cite{Giacinti:2012ar}, which also disagreed by 
an order of magnitude with the extrapolation of the diffusion coefficient 
phenomenologically determined from the ratio of secondary to primary nuclei.
\end{enumerate}
Thus our model is based on relatively small values of the coherence length
and the energy density in turbulent and regular magnetic fields.
The first assumption is supported by a number of observational studies
which derived limits  on the coherence length in the Galactic disk of
order 10\,pc~\cite{Iacobelli:2013fqa}. The second assumption appears
more contrived, since the required reduction factor $\beta\ap 1/8$ is 
relatively large. However, we note that the recent study~\cite{Beck:2014pma} 
suggests to rescale the isotropic turbulent field of the JF model by a 
factor 0.3, while it still predicts Faraday rotation measurements at low
galactic latitudes that are a factor two too large. Futhermore, a non-uniform  
density of electron in the Milky Way may lead to an over-estimate of the 
turbulent Galactic magnetic field.  Additionally, one
should be aware that several oversimplifications in our analysis may 
lead to a somewhat too large value of $\beta$: For instance, we have not 
properly accounted for a possible anisotropy in the turbulent magnetic field
or a spiral structure of CR sources in the Milky Way.

Let us note also that the weakness of the turbulent GMF in the
``escape model'' would have important consequence for the search of
UHECR sources: UHECRs from a single source would be mainly deflected by 
the regular component   of the GMF, while the spread  of their arrival
directions due to the turbulent GMF should be small. As a result,
the search for UHECR sources, at least in the case of protons or light nuclei, 
should be easier than thought before. Even for heavier nuclei, the
deflections in the regular field of the Galaxy can be traced back in those
patches of the  sky with small turbulent fields~\cite{Giacinti:2011uj}.
Weaker magnetic fields will also simplify the search of nuclei sources using
the methods discussed in Refs.~\cite{Giacinti:2009fy,Giacinti:2010dk}.
 Thus the results of this 
work are an additional motivation for future searches of UHECR sources,
performed by future all-sky  missions as e.g.\ JEM-EUSO~\cite{JE} and 
KLYPVE. ~\cite{KL}.

\subsection{Contribution of starbust galaxies}

It is natural to apply the 'escape model' to other normal galaxies. In 
particular, this model suggests that the CR knee in starburst galaxies
is shifted by two orders of magnitude to higher energies~\cite{KO14},
since the observed magnetic fields of these galaxies are a factor $\sim 100$
larger than in the Milky Way~\cite{Lacki:2013ry}. Therefore, the 
extragalactic CR flux in the intermediate energy region up to ankle 
should be composed mainly of CRs accelerated in starbust galaxies. 
The ankle is then interpreted as the transition to another extragalactic 
source class, as e.g.\ active galactic nuclei or gamma-ray bursts.

The flux of CRs escaping from starburst galaxies has a low-energy cutoff, 
when the interaction probability of CR nuclei on gas becomes of order one 
or the diffusion time in the intergalactic magnetic fields becomes comparable
to the Hubble time. The magnetic horizon can be
approximated at $E_{\rm cr}\lsim E\lsim 10^{18}$\,eV by~\cite{Aloisio:2004fz} 
\be
 r_{\rm hor}^2 = \int_0^{t_0}  \d t \:D(E(t)) 
               = \int_{E_0}^{E} \frac{\d\!E'}{\beta}\: D(E')
                \ap \frac{cl_c}{H_0} \: \left(\frac{E}{E_{\rm cr}}\right)^2 ,
\ee
where $H_0$ is the Hubble constant and the critical energy $E_{\rm cr}$ is 
defined by $R_{\rm L}(E_{\rm cr})=l_{\rm c}$.

Neither the strength nor the coherence length of the
intergalactic magnetic field are well known. Lower limits
on space-filling intergalactic magnetic field are around $10^{-17}$~\cite{low},
while the upper limit is given by $B\ap 0.1$\,nG~\cite{up}. Using for 
illustration $B = 10^{-12}$\,G and $l_{\rm c}=$\,Mpc, the size of the magnetic 
horizon at $E = 10^{16}$\,eV is for protons $r_{\rm hor} \sim 100$\,Mpc 
and becomes comparable to the Hubble radius at $E = 3\times  10^{16}$\,eV.
Since the magnetic horizon of nuclei is smaller, a further prediction
of this scenario is that the extragalactic flux is initially purely
composed of protons, becoming at higher energies heavier and more
similar to the Galactic composition. This is in line with our finding
of an extragalactic {\em proton\/} contribution above $3\times 10^{16}$\,eV,
cf.\ Fig.~\ref{Fig_exprot}, and a proton dominated extragalactic flux
deduced from the $\ln(A)$ data.

The low-energy cutoff due to interactions of CR nuclei on gas can be
estimated by rescaling Fig.~\ref{figX1}. For magnetic fields a factor
100 larger, significant attenuation of the CR nuclei sets in below
$10^{16}$\,eV. Thus we assume that the CR flux reaching the Milky Way
from starburst galaxies is not affected strongly by interactions or 
magnetic confinement in the for us interesting range  
$E\gsim 3\times  10^{16}$\,eV. Detailed calculations of the diffuse
CR flux from starburst galaxies will be presented in Ref.~\cite{preparation2}.

\subsection{Connection to diffuse neutrino and $\gamma$-ray fluxes}

The recent discovery~\cite{Aartsen:2014gkd,Aartsen:2014muf} of astrophysical 
neutrinos with energies $E>10$\,TeV  by 
the IceCube experiment opened a new field---high energy neutrino astrophysics. 
The measured astrophysical neutrinos events are not completely uniformly 
distributed over the sky, but have an over-density towards the Galactic plane
and a region close to the Galactic center. In Ref.~\cite{Neronov:2013lza}, 
it was suggested that these neutrinos are secondaries from CR interactions 
in the central part of our Galaxy. Later it was shown that the observed 
neutrino spectrum, which follows the power-law $1/E^{2.45}$~\cite{Aartsen:2014muf}, has the same 
slope and normalization as the all-sky gamma-ray spectrum measured by 
the Fermi-LAT experiment~\cite{FermiLAT} at lower energies~\cite{Neronov:2014uma}. 
It was suggested that both spectra are dominated by hadronic interactions of 
CRs in our Galaxy. In this case, the CR spectrum in the central part of the
Milky Way should be consistent with a $1/E^{2.5}$ power-law, which in turn 
agrees with the slope of the nuclei spectrum derived in this work. Taking into 
account the change of the power-law exponent by $1/3$ in the case of 
Kolmogorov turbulence, such a spectrum is consistent with the slope 
$1/E^{2.2}$ suggested by acceleration models. 	

An exception is the locally measured proton spectrum, which has the spectrum 
$1/E^{2.7}$. As argued in Ref.~\cite{Neronov:2014uma}, such a spectrum 
could by caused 'recent' (i.e.\ within $\sim 10$\,Myr) variation of 
the local CR flux due to a nearby source. Such an event might be connected
to the creation of the  Gould belt of molecular clouds. The aged proton 
spectrum of such source would be softer  than the average spectrum of 
Galactic ``sea'' CRs, while CR nuclei have been spallated except of at
very low rigidities. Finally, we note that the main contribution to the
observed amplitude of the dipole anisotropy at $E\lsim 10^{14}$\,eV could
be caused by this source, exceeding thereby the $1/E^{1/3}$ low-energy 
continuation of our estimate presented in Fig.~\ref{figd}~\cite{preparation1}. 

The flux of astrophysical neutrinos contains a significant fraction of 
neutrinos outside the Galactic plane, which should have an extragalactic 
origin. In the framework of the present model, these extragalactic neutrinos
should be created by CR interaction in normal and starburst Galaxies. 
As discussed in Ref.~\cite{Neronov:2014uma}, they can explain both 
a significant part of the diffuse gamma-ray background and of
the IceCube signal outside the Galactic plane~\cite{preparation2}.

\subsection{Source rate and the slope of the injection spectra}

In our model, we use as average injection spectrum of CRs a constant
power-law $dN/dE\sim E^{-\alpha_i}$ for each group of CR nuclei over the 
rigidity range 200 to $10^8$\,GV. This suggests that a single source class 
accelerates the CRs observed in this energy range. These sources should
either all accelerate to $\Rm\simeq 1\times 10^{17}$\,V, or their maximal
rigidities should follow approximately a power-law 
distribution~\cite{Kachelriess:2005xh}.
Since the required maximal energy in our model is high, one expects
that only a subset of all Galactic CR accelerator is responsible for the
CRs in this energy range.  This is in line with our determination of
the source rate, 0.4/century, that is a factor ten lower than the 
usually assumed SN rate.

Because of the large leverage in our fits, the resulting constraints on 
the exponents $\alpha_i$ are much tighter than considering only, e.g., 
CREAM data: Rather steep power-laws with, e.g.\ $\alpha_p\simeq 2.65$ 
for the observed proton spectrum, which cannot be excluded by CREAM data 
alone are incompatible adding KASCADE and KASCADE-Grande data.

\section{Conclusions}
\label{Conclusions}

We have shown that the knee can be entirely explained by energy-dependent 
CR leakage from the Milky Way, with an excellent fit to {\it all\/} 
existing data from $E/Z\sim 300$\,GeV to 100\,PeV. 
In particular, all deviations from a single power-law behavior that are 
observed in the CR intensity of individual CR groups in the energy range 
$E/Z\sim 200$\,GeV up to 100\,PeV are consistently explained by 
rigidity-dependent CR escape.  This model requires small coherence lengths 
of the turbulent field and relatively small turbulent magnetic fields. 
If these two conditions are fullfilled, then
the CR escape time $\tau_{\rm esc}(E)$ exhibits a knee-like structure around 
$E/Z={\rm few}\times 10^{15}$\,eV together with a recovery around
$E/Z\simeq 10^{16}$\,eV.

We have determined the maximal rigidity $\Rm=E_{\max, \rm Fe}/(26e)\sim 10^{17}$\,V 
to which Galactic  CR sources are able to accelerate CRs by identifying it 
with the energy where $\ln(A)$ derived from PAO measurements stops to 
decrease. The resulting flux ratio of Galactic and extragalactic sources is 
in our model 1:1 at $E_\ast\ap 2\times 10^{17}$\,eV, dropping to 0:1  at 
$2\times 10^{18}$\,eV. 
The extragalactic CR flux in the intermediate energy region up to ankle 
should be composed mainly of CRs accelerated in starbust galaxies.
Since the transition from Galactic to extragalactic CRs happens in 
this model at rather low energies, the estimated CR dipole anisotropy is
consistent within uncertainties with upper limits in the energy range 
$10^{17}-10^{18}$\,eV, while it reproduces the measurements at lower
energies from EAS-TOP and IceCube. 
The dipole phase is expected to change between 
$1\times 10^{17}$ and $3\times 10^{18}$\,eV, i.e.\ the energy range of the
transition from Galactic to extragalactic CRs. Such a behavior corresponds 
to the one observed, providing thus additional
evidence for a transition from Galactic to extragalactic CRs
in this energy region.

\acknowledgments

We would like to thank  Andreas Haungs for discussions about the 
KASCADE and KASCADE-Grande data and Michael Unger for sending us data 
files of the $\ln(A)$ values derived in~\cite{Kampert:2012mx}.
MK thanks the Theory Group at APC for hospitality.
GG acknowledges funding from the European Research Council under the European 
Community's Seventh Framework Programme (FP7/$2007-2013$)/ERC grant 
agreement no. 247039.
The work of DS was supported in part by grant RFBR \# 13-02-12175-ofi-m.


\end{document}